\documentclass[twocolumn,showpacs,preprintnumbers,amsmath,amssymb]{revtex4}

\usepackage{graphicx}
\usepackage{dcolumn}
\usepackage{bm}

\begin{document}

\preprint{}
\renewcommand{\thefootnote}{\fnsymbol{footnote}}

\title{Dissociation and Decay of Ultra-cold Sodium Molecules}

\author{T. Mukaiyama, J. R. Abo-Shaeer, K. Xu, J. K. Chin, and W. Ketterle}

\affiliation{Department of Physics, MIT-Harvard Center for
Ultracold Atoms, and Research Laboratory of Electronics,
Massachusetts Institute of Technology, Cambridge, Massachusetts
02139}

\date{\today}

\begin{abstract}
The dissociation of ultracold molecules is studied by ramping an
external magnetic field through a Feshbach resonance. The observed
dissociation energy shows non-linear dependence on the ramp speed
and directly yields the strength of the atom-molecule coupling. In
addition, inelastic molecule-molecule and molecule-atom collisions
are characterized.
\end{abstract}

\maketitle

Recently, it has become possible to create ultracold molecular
gases from precooled atomic
samples~\cite{Wynar,Donley,Regal,Jochim,Cubizolles,Strecker,Durr,Chin,Xu,Greiner,Jochim2}.
Extending the ultralow temperature regime from atoms to molecules
is an important step towards controlling the motion of more
complicated objects. The complex structure of molecules may lead
to new scientific opportunities, including the search for a
permanent electric dipole moment, with sensitivity much higher
than for heavy atoms~\cite{Hudson}, and the realization of quantum
fluids of bosons and fermions with anisotropic
interactions~\cite{Baranov}. Furthermore, stable mixtures of
atomic and molecular condensates are predicted to show coherent
stimulation of atom-molecule or molecule-atom conversion,
constituting quantum-coherent chemistry \cite{Heinzen}.

To date, all realizations of ultracold molecules have bypassed the
need for direct cooling of the molecules, which is difficult due
to the complicated ro-vibrational structure. Rather, molecules
were formed from ultracold atoms using Feshbach
resonances~\cite{Donley,Regal,Jochim,Cubizolles,Strecker,Durr,Chin,Xu,Greiner,Jochim2},
where a highly-vibrational excited molecular state is magnetically
tuned into resonance with a pair of colliding atoms.

\begin{figure}[t]
\begin{center}
\includegraphics[width=\linewidth]{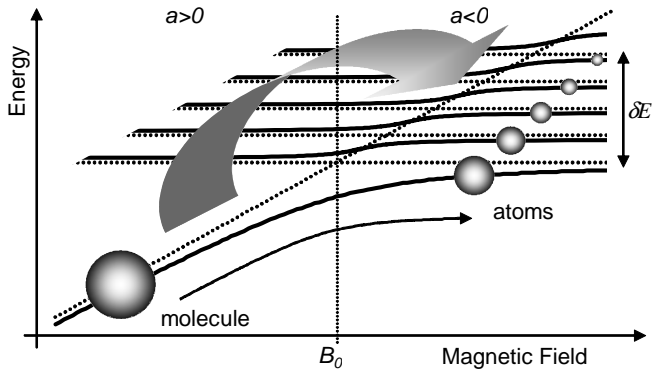}
\end{center}
\caption{Schematic diagram of energy levels for molecules and
atoms. The diabatic energy levels are shown as dashed lines. The
adiabatic curves (solid lines) include the atom-molecule coupling.
When the magnetic field is swept from positive to negative
scattering length, dissociated molecules end up in one or several
atomic states, depending on the ramp rate of the magnetic field.
The spheres represent the distribution of the population before
and after the ramp.} \label{fig:schematic}
\end{figure}

In this paper, we study the dissociation and decay of such highly
excited molecules. Figure~\ref{fig:schematic} shows the relevant
energy levels. For magnetic fields above the Feshbach resonance,
the molecular state crosses the free atomic states, shown here as
discrete states in a finite quantization volume. The interaction
between atoms and molecules turns these crossing into
anti-crossings. When the magnetic field is swept very slowly
through the resonance, the molecules will follow the adiabatic
curve and end up in the lowest energy state of the atoms. For
faster ramps, the molecular populations will partially cross some
of the low-lying states, and the dissociation products will
populate several atomic states. The stronger the coupling between
the molecular state and the atomic states, the faster the
molecules dissociate and the smaller the energy release in the
dissociation. Observing the atom-molecule coupling in {\it
one-body} decay (dissociation) is a new method to experimentally
determine the strength of a Feshbach resonance. Previous
measurements used {\it two-} or {\it three-body} processes to
characterize the Feshbach resonance and therefore required
accurate knowledge of the atomic density distribution.

Collisional properties of the molecules were also studied.
Inelastic collisions limit both the production of molecules and
their lifetime. We observed loss of molecules by collisions both
with atoms and other molecules. These two processes were studied
separately because we could produce atom-molecule mixtures, as
well as pure molecular samples, by separating atoms and molecules
with short pulses of laser light~\cite{Xu}.


To generate molecules, sodium condensates in the
$|F,m_F\rangle=|1,-1\rangle$ state were prepared in an optical
dipole trap. The trap frequencies of 290~Hz in the radial
direction and 2.2~Hz in the axial direction yielded typical
densities of $1.7 \times 10^{14}$~cm$^{-3}$ for 5 million atoms.
Atoms were then spin-flipped using an adiabatic radio frequency
sweep to the $|1,1\rangle$ state, where a 1~G wide Feshbach
resonance exists at 907~G~\cite{Inouye}.

The magnetic field sequence used to create and detect Na$_2$
molecules was identical to our previous work~\cite{Xu}. Briefly,
the axial magnetic field was ramped to 903~G in 100~ms using two
pairs of bias coils. In order to prepare the condensate on the
negative scattering length side of the resonance, the field was
stepped up to 913~G as quickly as possible ($\sim$~1~$\mu$s) to
jump through the resonance with minimal atom loss. The field was
then ramped back down to 903~G in 50~$\mu$s to form molecules. In
order to remove non-paired atoms from the trap, the sample was
irradiated with a 20~$\mu$s pulse of resonant light. Because 903~G
is far from the Feshbach resonance, the mixing between atomic and
molecular states was small, and therefore molecules were
transparent to this ``blast'' pulse. By ramping the field back to
913~G molecules were converted back to atoms. Absorption images
were taken at high fields (either at 903~G or 913~G) after 10 to
17~ms ballistic expansion, with the imaging light incident along
the axial direction of the condensate.

\begin{figure}[b]
\begin{center}
\includegraphics[width=\linewidth]{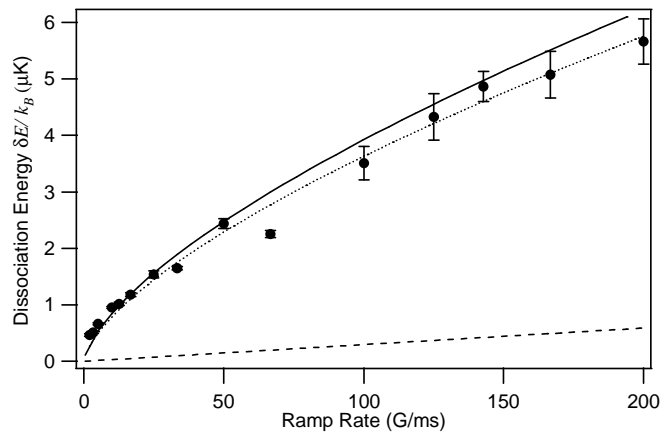}
\end{center}
\caption{Dissociation energy of molecules as a function of
magnetic field ramp rate. The dashed line represents the linear
relation described in ref.~\cite{Abeelen}, the solid line shows
the result of our theory with no free parameters (using a
theoretical value for $\Delta B=0.98$ G), and the dotted line
shows a curve with $\Delta B$ as a fitting parameter.}
\label{fig:dissociation}
\end{figure}

To study the momentum distribution of the back-converted atoms,
the magnetic field was ramped up immediately after turning off the
optical trap, or for a reference, at the end of the ballistic
expansion. The difference between the energies of ballistic
expansion is the released dissociation energy. Energies were
obtained from the rms width of the cloud $\langle x^2 \rangle$ as
$E=3m\langle x^2 \rangle/2t^2$, where $t$ is the ballistic
expansion time, and $m$ is the atomic mass.
Figure~\ref{fig:dissociation} shows that faster field ramps
created hotter atoms.

An earlier theoretical treatment assumes a constant
predissociation lifetime of the molecules and predicts a linear
relation between dissociation energy and field ramp
rate~\cite{Abeelen}. This theory predicts a much faster
dissociation (and therefore smaller dissociation energy) than was
observed. Furthermore, our data shows a non-linear dependence.
Linear behavior would be expected if the lifetime of the molecules
was independent of the energy $\epsilon$ from the dissociation
threshold. The fact that the slope becomes smaller for increasing
ramp rate indicates that the lifetime of molecules decreases with
the ramp rate. As we will show, this can be explained by an
increase of the density of atomic states, leading to a
$\sqrt{\epsilon}$ dependence of the molecular decay rate (Wigner
threshold law~\cite{Wigner}).

 The decay rate $\Gamma(\epsilon)$ follows from Fermi's golden
rule as $\hbar \Gamma(\epsilon)=2\pi|V_{ma}(\epsilon)|^2
D(\epsilon)$~\cite{Mies}, where $V_{ma}$ is the matrix element
between atomic and molecular states, which to leading order is
independent of $\epsilon$. The density of states $D(\epsilon)$ is
given by
\begin{equation}
D(\epsilon)=\frac{V}{(2\pi)^2}
\left(\frac{m}{\hbar^2}\right)^{3/2}\epsilon^{1/2},
\label{eq:density_states}
\end{equation}
where $V$ is the quantization volume for free atomic states.

An expression for the matrix element $V_{ma}$ is obtained by
comparing the energy shift near a {Feshbach} resonance with
second-order perturbation theory. Assuming two atoms in a volume
$V$, the energy shift of the low-lying continuum states due to the
coupling with a bound molecular state is
\begin{equation}
\delta (\epsilon)=\frac{|V_{ma}|^2}{ \epsilon}
=\frac{|V_{ma}|^2}{\Delta \mu (B-B_0) }, \label{eq:energyshift1}
\end{equation}
where $\Delta\mu$ is the difference between atomic and molecular
magnetic moments, $B$ the applied magnetic field, and $B_0$ the
position of the Feshbach resonance.

The energy shift can also be expressed in terms of the mean field
energy $4\pi \hbar^2a/mV$, where $a=a_{bg} \Delta B / (B-B_0)$ is
the scattering length near the Feshbach resonance ($a_{bg}$ is the
background scattering length and $\Delta B$ is the resonance
width~\cite{Timmermans}):
\begin{equation}
\delta (\epsilon) =\frac{4\pi \hbar^2}{mV} \frac{a_{bg} \Delta B
}{B-B_0}.\label{eq:energyshift2}
\end{equation}
Comparing eq.~(\ref{eq:energyshift1}) and
eq.~(\ref{eq:energyshift2}) yields
\begin{equation}
|V_{ma}|^2=\frac{4\pi\hbar^2}{mV}a_{bg}\Delta \mu \Delta B.
\label{eq:energyshift3}
\end{equation}

If the entire population is initially in the molecular state, the
fraction of molecules, $m(\epsilon)$, at energy $\epsilon$ follows
the rate equation,
\begin{eqnarray}
n\frac{dm(\epsilon)}{d\epsilon}&=&
\frac{dm(\epsilon)}{dt}\left(\frac{d\epsilon}{dt}\right)^{-1}=
\Gamma(\epsilon)m(\epsilon)\left(\frac{d\epsilon}{dt}\right)^{-1}
\\
&=&\frac{2\pi |V_{ma}(\epsilon)|^2 D(\epsilon)}{\hbar \Delta \mu
|\dot{B}|}m(\epsilon). \label{eq:rateequ}
\end{eqnarray}

Using Eqs.~(\ref{eq:density_states}) and (\ref{eq:energyshift3}),
we solve the differential equation for $m(\epsilon)$
\begin{eqnarray}
m(\epsilon)&=&e^{-\frac{2}{3}C\epsilon^{3/2}}, \\
C&=& \frac{2\Delta B}{\hbar \dot{B}}\sqrt{\frac{m
a_{bg}^2}{\hbar^2}} \nonumber.
\end{eqnarray}

In the lab frame, the atoms have kinetic energy $\epsilon/2$ and
therefore the average energy of an atom after dissociation is
\begin{equation} \delta E =
\int_0^{\infty}\frac{\epsilon}{2}\left(-dm(\epsilon)\right) =
0.591 \left(\sqrt{\frac{\hbar^2}{m a_{bg}^2}}\frac{\hbar
\dot{B}}{2\Delta B} \right)^{2/3}.
\end{equation}

\begin{figure}[t]
\begin{center}
\includegraphics[width=\linewidth]{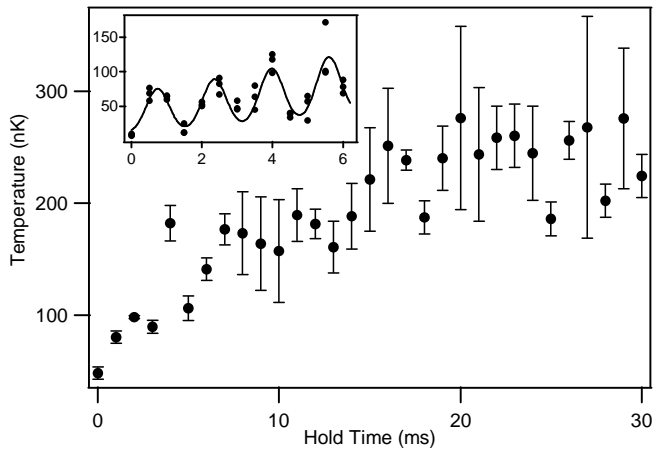}
\end{center}
\caption{Temperature of the molecular cloud. After 15~ms, the
temperature saturates at $\sim$~250~nK. Error bars represent the
statistical error (standard deviation). The inset shows finer
resolution data for holding times up to 6 ms. The solid line is a
guide to the eye.} \label{fig:temphold}
\end{figure}

Using theoretical values $a_{bg}=3.4$~nm, $\Delta \mu /h=
3.65\times 1.4$~MHz/G, and $\Delta B=0.98$~G
\cite{Yurovsky43611,Yurovsky43605}, our parameter-free prediction
(solid line in Fig.~\ref{fig:dissociation}) shows good agreement
with the experimental data. Alternatively, we may regard the width
of the Feshbach resonance as a free parameter to obtain a fitted
value of $\Delta B=1.07 \pm 0.02$~G (dotted line in
Fig.~\ref{fig:dissociation}). Compared to previous mean-field
measurements~\cite{Inouye,Stenger}, our determination of the
resonance width is more accurate and free from systematic errors
associated with the determination of atomic densities.

Further experiments with ultracold sodium molecules will
critically depend on their collision properties. Therefore we also
studied heating and inelastic collision processes. As shown in
Fig.~\ref{fig:temphold}, we observed monotonic heating of the pure
molecular sample over $\sim$~30~ms. In addition, we observed short
timescale oscillations (6~ms) in the fitted temperature (inset of
Fig.~\ref{fig:temphold}). Such breathing oscillations were excited
because the molecules were formed over the volume of the atomic
condensate, which was larger than the equilibrium volume for the
molecules. The absence of damping implies a collision time of at
least 6~ms, or a molecular scattering length smaller than 17~nm
(obtained using the expression for the collision rate $8\pi a^2
v_{th}n_m$ where $v_{th}$ is the thermal velocity). It is unclear
whether the oscillation disappeared due to collisions or limited
signal-to-noise ratio.

The temperature of the molecular cloud saturated at $\sim$~250~nK
after 15~ms. A possible explanation is the balance between heating
due to inelastic molecular decay and the evaporative cooling
caused by the finite trap depth (1.7~$\mu$K). This would imply a
collision time of 15~ms. However, we have no clear evidence that
thermalization has occurred. Clearly, further studies of elastic
collisions between ultracold molecules are necessary.

Molecules formed via Feshbach resonances are created in high
vibrational states. Therefore, one expects vibrational relaxation
to be a strong, inelastic decay mechanism. Vibrational energy
spacings are much larger than the trap depth, leading to loss of
molecules from the trap.

\begin{figure}[b]
\begin{center}
\includegraphics[width=\linewidth]{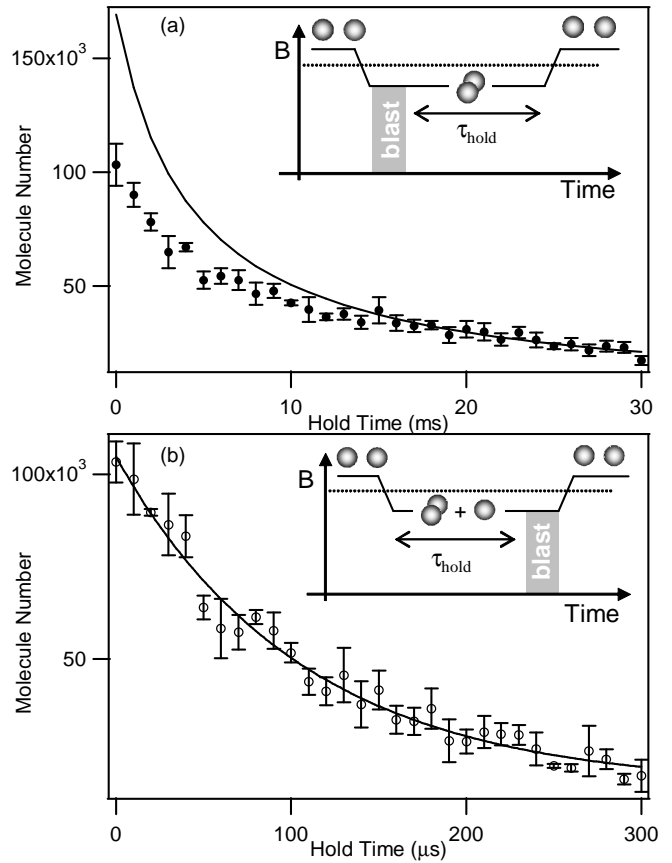}
\end{center}
\caption{Decay of ultracold molecules trapped alone (a) or
together with atoms (b). The solid lines in (a) and (b) are fits
of eq.(\ref{eq:mol}) and (\ref{eq:atom}) to data, which assume
vibrational relaxation in the collision of molecules (a) or
collisions between molecules and atoms (b). The insets illustrate
the experimental sequences.} \label{fig:decay}
\end{figure}

 Figure \ref{fig:decay}(a) shows the decay of a pure molecular
sample. The decay was analyzed with the rate equation
\begin{equation}
\frac{\dot{N}_{m}}{N_{m}}=-K_{mm} n_{m}.\label{eq:mol}
\end{equation}
Here $n_{m}$ is the density of the molecules, and $K_{mm}$ is the
molecule-molecule collision rate coefficient.  Because of the
changing size and temperature of the molecular cloud during the
first $\sim$~15~ms (Fig.~\ref{fig:temphold}), we only fit data
points at later times, assuming a thermal equilibrium volume for
the molecules. The decay at earlier times is slower, consistent
with a larger molecular cloud. The fit yields a molecule-molecule
collision coefficient of $K_{mm} \sim 5.1 \times
10^{-11}$~cm$^{3}$/s, about 2 orders of magnitude larger than the
typical values reported for fermions~\cite{Jochim,Cubizolles}.

\begin{figure}[t]
\begin{center}
\includegraphics[width=\linewidth]{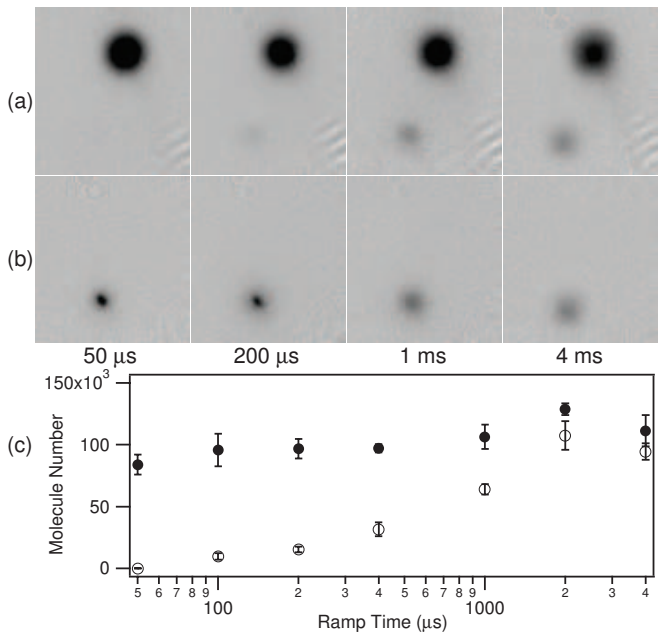}
\end{center}
\caption{Conversion of atoms to molecules for various ramp times.
During a given time, the magnetic field was swept by 10~G.
Figures~(a) and (b) show absorption images taken after 14~ms TOF.
The molecules (bottom) were radially separated from the atoms
(top) by a field gradient of 2.8~G/cm. The molecules were
converted back to atoms only 0.5~ms before imaging by ramping the
magnetic field back across the Feshbach resonance. This time was
chosen to be long enough for any transient field fluctuations to
damp out, but short enough such that the size of the imaged cloud
reflected the molecular temperature, not the dissociation energy.
(a) The atoms remained in the trap. (b) The atoms were removed by
a resonant laser pulse immediately after the magnetic field ramp.
(c) Number of molecules as a function of ramp time for (a) (open
circles) and (b) (closed circles).} \label{fig:pictures}
\end{figure}

Inelastic collisions between molecules and atoms were also
observed by keeping atoms in the trap (Fig.~\ref{fig:decay}(b)).
The decay was analyzed assuming that the loss of molecules
occurred mainly due to collisions with atoms, resulting in an
exponential decay:
\begin{equation}
\frac{\dot{N}_{m}}{N_{m}}=-K_{am} n_{a}.\label{eq:atom}
\end{equation}
Here $N_{m}$ is the number of the molecules, $n_{a}$ is the
density of atoms, and $K_{am}$ is the atom-molecule collision rate
coefficient. From the fit, we extract a lifetime of 106~$\mu$s and
a rate coefficient $K_{am} \sim 5.5 \times 10^{-11}$~cm$^{3}$/s,
which agrees well with theoretical
predictions~\cite{Yurovsky43605, Yurovsky43611}.

The inelastic losses determine the maximum conversion efficiency
from atoms to molecules. For an adiabatic ramp, one expects close
to 100\% conversion efficiency. Indeed, in experiments with
fermionic atoms, efficiencies up to 85\% have been
observed~\cite{Cubizolles}. Figure~\ref{fig:pictures} shows the
results for magnetic field ramps of different durations. The two
sets of images show that applying the blast pulse dramatically
improved the molecular number and temperature. Without it, a
slower ramp time (4~ms) appeared to be more favorable for molecule
formation (open circles in Fig.~\ref{fig:pictures}(c)). No
molecules were observed for a 50~$\mu$s ramp time. However, with
the blast pulse, nearly the same number of molecules was obtained
for all ramp times between 50~$\mu$s to 4~ms (closed circles in
Fig.~\ref{fig:pictures}(c)).

We interpret our data as the interplay of two competing processes.
The adiabatic condition requires a relatively slow field ramp for
efficient conversion. However, this means that the atoms and
molecules spend more time near or at the Feshbach resonance, where
inelastic collision rates are enhanced. In contrast to
Fig.\ref{fig:pictures}(b), the absence of molecular signal in
Fig.~\ref{fig:pictures}(a) for 50~$\mu$s ramp time reflects that
the atomic density reduction due to the mean-field expansion is
too slow for the molecules to survive the inelastic collisions
with the atoms.

In conclusion, we observed a Wigner threshold behavior in the
dissociation of ultracold molecules. We were able to characterize
a Feshbach resonance using a one-body decay (dissociation)
process. The rapid decay of the molecules due to collisions with
atoms and other molecules imposes a severe limit to further
evaporative cooling for bosons. This also explains the low
conversion efficiency ($\sim$~4\%), in contrast to recent
experiments with fermions.

The authors would like to thank D. E. Miller for experimental
assistance and A. E. Leanhardt, M. Saba and D. Schneble for their
critical reading of the manuscript. We also thank P. S. Julienne,
B. J. Verhaar, and V. A. Yurovsky for useful discussion. This
research is supported by NSF, ONR, ARO and NASA.

\end{document}